\documentclass[reprint,amsmath,amssymb,aps,prl,longbibliography]{revtex4-1}
\usepackage[english]{babel}

\preprint{APS/123-QED}
\usepackage{graphicx}
\usepackage{bm}
\usepackage{physics}

\newcommand{\TKE}{T\!K\!E}
\newcommand{\R}{\textrm{Re}}

\begin{document}

\title{Self-Replication of Turbulent Puffs: On the edge between chaotic saddles}

\author{Anton Svirsky$^{1}$ , Tobias Grafke$^{2}$ and Anna Frishman$^1$}

\email{frishman@technion.ac.il}

\affiliation{$^1$Physics Department, Technion Israel Institute of Technology, 32000 Haifa, Israel}
\affiliation{$^2$Mathematics Institute, University of Warwick, Coventry CV4 7AL, United Kingdom}

\begin{abstract}

    Pipe flow is a canonical example where turbulence first appears intermittently in space and time, taking the form of localized structures termed puffs. Turbulence spreads via puff self-replication, which must out-compete puff decays to sustain it. Here we study the self-replication process, a transition from one to two puffs, using direct numerical simulations. We identify an edge state on the phase space boundary between the two states, demonstrate that it mediates the transition, and show that self-replication follows a previously proposed mechanism, with the edge state as its tipping point.

\end{abstract}

\date{\today}

\maketitle

    The route to turbulence in wall-bounded shear flows has puzzled researchers for over a century. Understanding this transition is not only a fundamental challenge in fluid mechanics but also a problem of practical importance, as turbulence significantly impacts drag, mixing, and energy efficiency. A canonical example is the transition to turbulence in pipe flow, which has remained an active area of research ever since Osborne Reynolds' seminal study \cite{reynolds_experimental_1883}.

    In pipe flow, the transition to turbulence depends solely on the Reynolds number ($\R$), which quantifies the ratio of shear to dissipation rates. Laminar pipe flow is linearly stable up to at least $\R=10^7$ \cite{mullin_iutam_2005}, so turbulence requires a finite-amplitude perturbation to be excited. The transition is thus governed by non-linear processes, and is termed subcritical \cite{drazin_introduction_2002}. Sustained turbulence emerges above a critical Reynolds number $\R_c \approx 2040$ \cite{avila_onset_2011}, below which turbulent structures eventually decay. 

    The subcritical nature of the flow implies that turbulence can also be intermittent in space. Indeed, in the range $1800 \lesssim \R \lesssim 2300$, turbulence takes the form of localized patches, termed \emph{puffs}, surrounded by laminar flow  \cite{wygnanski_transition_1973, wygnanski_transition_1975, nishi_laminar-turbulent_2008, van_doorne_flow_2009}. Then, two competing processes determine the fraction of turbulence occupying the pipe: puff self-replications (splits), which increase the turbulence fraction, and puff decays, which decrease it \cite{hof_scaling_2003, willis_critical_2007, barkley_theoretical_2016}. Both are memoryless processes with lifetimes that depend on $\R$.
    Above a critical $\R =\R_c$, where splits become more frequent than decays, a finite fraction of turbulence can be sustained \cite{avila_onset_2011, hof_finite_2006}. 
    
    The competition between self-replication and decay of turbulent structures is the generic route to sustained turbulence in subcritical wall-bounded flows, examples ranging from classical high $\Re$ flows \cite{manneville_transition_2016,hof_directed_2023}, to active turbulence \cite{doostmohammadi_onset_2017}.
Yet the mechanism behind the self-replication process remains poorly understood. Here, we make significant progress in filling this knowledge gap for pipe flow: we identify a key dynamical state at the phase space boundary between one and two puffs, and demonstrate that splitting events follow a generic transition path, with this state serving as its "tipping point". The emergent transition path is consistent with a previously proposed splitting mechanism \cite{frishman_mechanism_2022}, providing its first direct confirmation.

    We consider an incompressible fluid (density $\rho=1$) with viscosity $\nu$, flowing in a pipe with circular cross-section of diameter $D$ and length $L$ with periodic boundary conditions in the stream-wise direction. The Reynolds number is defined as $\R= \overline{U}D/\nu$, where $\overline{U}$ is the cross-section mean flow velocity which is kept constant, enforcing constant mass flux. The equations are non-dimensionalized with velocity and length scales $\overline{U}$ and $D$ respectively. The Navier-Stokes equations are integrated using the \texttt{openpipeflow} code \cite{willis_openpipeflow_2017}, see end-matter for the parameters used.

       Below, to emphasize the difference from laminar flow, we use the perturbation velocity $\bm{u}$ defined such that $\bm{u} = 0$ for laminar flow.
   The presence of turbulence is quantified by
    \begin{equation}
        \label{eq:q_var}
        q(z,t) = \frac{4}{\pi D^2}
        \int_0^{2\pi} \int_0^{D/2} r \, \dd \theta \, \dd r \, (u_r^2 + u_\theta^2) \, ,
    \end{equation}
    where $q(z) > 0$ indicates the presence of turbulence at axial position $z$, since the radial and azimuthal velocity components vanish for laminar flow.
    
    \textbf{Dynamical systems picture} 
    To set the stage for the understanding of puff self-replication as a dynamical process, we first recall the analogous picture for puff decays. From a dynamical systems point of view  \cite{hopf_mathematical_1948,itano_dynamics_2001, manneville_transition_2016,schneider_turbulence_2007,eckhardt_turbulence_2007, avila_transition_2023, duguet_transition_2008, faisst_sensitive_2004, budanur_relative_2017}, each velocity field $s = \bm{u}(\bm{r})$ is a point in a (potentially infinite-dimensional) phase space $\Omega$---the set of all incompressible velocity fields satisfying the boundary conditions. The system evolves via the Navier-Stokes equations, which can be formally expressed as an autonomous dynamical system: $\dot{s} = F(s;\R)$, tracing a trajectory $s(t)$ in $\Omega$.

\begin{figure}[t] 
        \centering
        \includegraphics[width=\columnwidth]{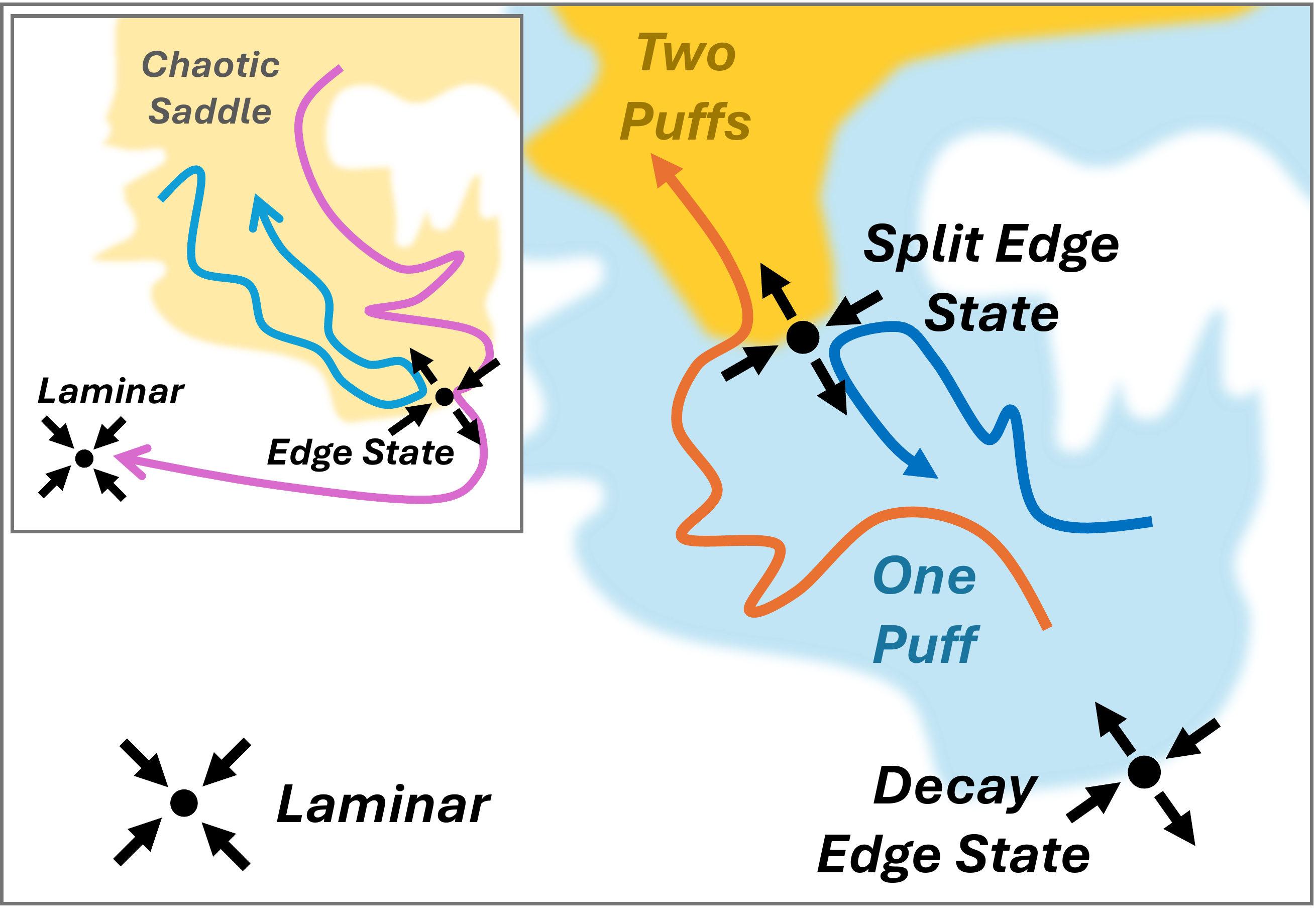}
        \caption{
        Illustration of one- and two-puff configurations as distinct chaotic saddles. An orange trajectory from the one-puff state reaches the split-edge state and transitions to the two-puff chaotic saddle via its unstable manifold, while a blue trajectory returns to the one-puff chaotic saddle. The inset depicts turbulence as a single chaotic saddle: the blue trajectory explores the saddle, whereas the pink trajectory hits the edge state and escapes along its unstable manifold, leading to decay.
        }\label{fig:phase_space_cartoon}
    \end{figure}

    Within this picture, the laminar state is a stable fixed point, while a puff is a trajectory near a chaotic, non-attracting set known as a chaotic saddle (since puffs are transient, they are not attractors). The laminar state and the puff state are separated by an effective phase space boundary, known as the \emph{edge of chaos} \cite{schneider_turbulence_2007,eckhardt_transition_1999,skufca_edge_2006,duguet_transition_2008}, defined as the boundary between initial conditions that quickly relaminarize and those that produce puffs first.
The edge of chaos is expected to form the stable manifold of a saddle set embedded within it, known as an \emph{edge state} (see inset of Fig.~\ref{fig:phase_space_cartoon}). This has been confirmed in direct numerical simulations (DNS) in short  \cite{eckhardt_transition_1999,duguet_relative_2008,schneider_turbulence_2007} and long \cite{mellibovsky_transition_2009,duguet_slug_2010,avila_streamwise-localized_2013} pipes.  Trajectories near the edge tend to evolve toward the edge state, and then either decay or become turbulent, depending on which ``side of the boundary'' they fall on \cite{budanur_heteroclinic_2017}. 
     
    From this point of view, the decay of a puff corresponds to an escape from a chaotic saddle \cite{avila_transition_2023, ott_chaos_2002}, explaining the exponential waiting times of the process. The edge of chaos is necessarily crossed during laminarization, and it is natural to expect this crossing to occur close to the edge state. It has indeed been demonstrated in experiments that puffs approach the edge state before decaying \cite{de_lozar_edge_2012}. For this reason, in the following we will refer to this edge state as the \emph{decay edge} state.
    
    Puff splitting, like decay, is a memoryless process with exponential waiting times, suggesting it too corresponds to an escape from a chaotic saddle. However, unlike decay, splitting requires a transition between two chaotic saddles: one with $n$ puffs and another with $n+1$. We focus on the fundamental one-to-two-puffs transition, treating the two states as distinct chaotic saddles separated by a phase-space boundary. Then, we expect puff self-replication, like decay, to be mediated by an edge state embedded in the phase-space boundary separating these two chaotic saddles, as illustrated in Fig.~\ref{fig:phase_space_cartoon}. We term this edge state the \emph{split edge} state. 

    \begin{figure*}[h!t]  
        \centering
        \includegraphics[width=2\columnwidth]{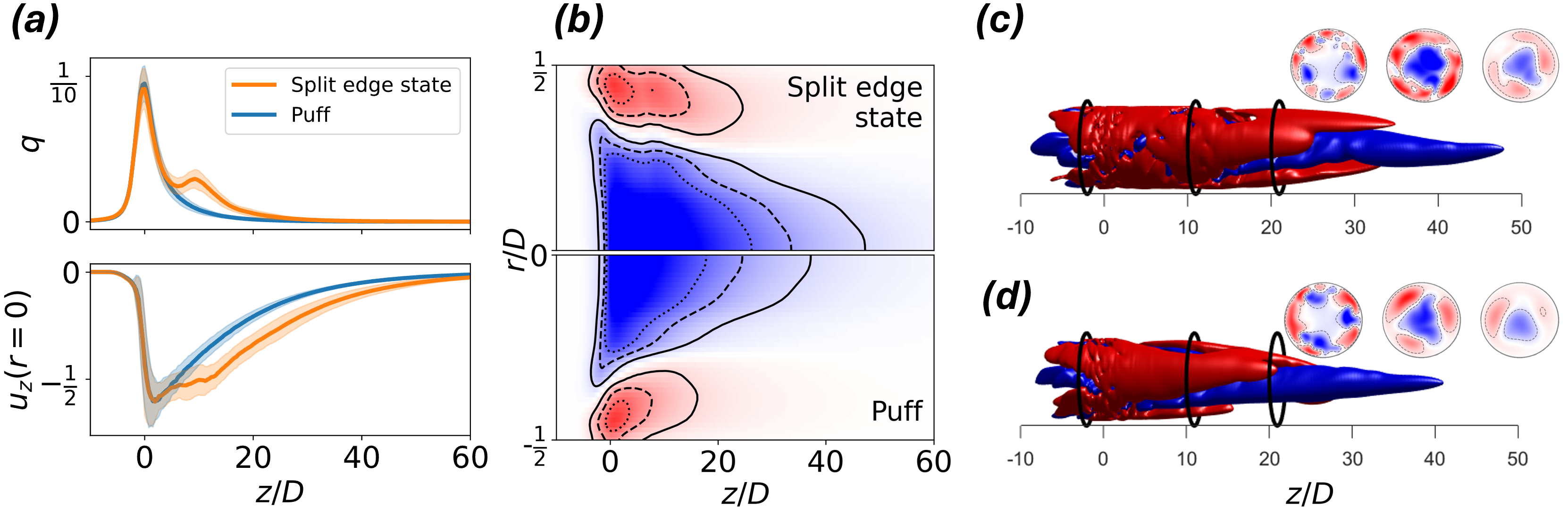}
        \caption{
        Comparison of flow profiles for the split edge and turbulent puff at $\R = 2200$. (a) Time-averaged turbulence level $q$ (Eq.~\eqref{eq:q_var}) and centerline perturbation velocity $u_z(r=0)$; shading shows one standard deviation. (b) Cross-section of the azimuthally and time-averaged axial velocity $u_z$ of the split edge with contours at $u_z = \pm 0.07$ (solid), $\pm 0.15$ (dashed), and $\pm 0.24$ (dotted). (c,d) Instantaneous $u_z$ for the edge state (c) and puff (d). Isosurfaces $u_z = 0.07$ (red) and $u_z = -0.07$ (blue) are shown with three cross-sections marked by black rings. In (b–d), blue, white, and red denote $u_z$ lower than, equal to, and higher than the laminar profile.\label{fig:Re2200_es}}  
    \end{figure*} 

    We now set out to confirm this picture. To identify the split edge state, we employ a bisection method similar to that used in studies of the edge of chaos \cite{skufca_edge_2006}: iteratively refining two nearby initial conditions that evolve toward the one- and two-puff states. Their trajectories remain near the stable manifold of the split edge state and diverge along its unstable one, and are thus expected to approach the edge state after an initial transient. Details of this edge-tracking algorithm are provided in the end-matter.

   
    \textbf{Split edge state} We apply the edge-tracking algorithm at $\R= 2200$, a regime where puff splitting can be observed in DNS within a reasonable time frame, allowing for explicit verification of the edge-states' relevance to the splitting process. After an initial transient of the algorithm, a unique spatially localized structure is obtained, which we refer to as the \emph{split edge}. We can estimate the maximal Lyapunov exponent for this edge state from the temporal divergence of the $L^2$ distance between two edge-bounding states, and find it to be approximately $\lambda = 0.48 \pm 0.04 \ [\overline{U}/D] $.
    
    In terms of its structure, the split edge closely resembles an elongated puff, with its upstream and downstream fronts exactly identical to those of the puff state (Fig.~\ref{fig:Re2200_es}), and its travel speed matching that of the puff. Furthermore, the azimuthal structure of the edge state is qualitatively similar to that of a puff, as illustrated by the instantaneous flow field in Fig.~\ref{fig:Re2200_es}(c,d).  
    The relative elongation of the split edge is due to the presence of a (relatively) homogeneous turbulent core immediately after the upstream front. This core extends over a length of approximately $8 D$, where both the centerline velocity $u_z(r=0)$ and turbulence level $q$ remain nearly constant, as shown in Fig.~\ref{fig:Re2200_es}(a,b). 
    
    The split edge thus resembles a short slug -- a turbulent structure similar to a puff but with a homogeneous turbulent core, known to replace puffs at $\R >2300$ where it is an expanding structure. 
    Notably, this form of the split edge state agrees with the prediction put forth in~\cite{frishman_mechanism_2022}. Correspondingly, the $q,u_z$ profiles in Fig.~\ref{fig:Re2200_es}(a,b) closely resemble those of the split edge found in the Barkley model (a phenomenological model of pipe flow~\cite{barkley_theoretical_2016}) using a similar bisection method~\cite{frishman_mechanism_2022}. 
    
\textbf{Relevance for splitting events}
    While the split edge state lies on the boundary between the one-puff and two-puff chaotic saddles, splitting trajectories do not necessarily approach it. To test if they do, we analyze $N = 9$ naturally occurring splitting events from DNS at $\R = 2200$, treating each as a trajectory in the high-dimensional phase space $\Omega$. To quantify the transition paths and their proximity to the split edge state, we seek a low-dimensional phase-space projection that best separates the one-puff, two-puff, and split edge states. A data-driven, physics-agnostic choice for such a reduction can be obtained by employing principal component analysis (PCA) \cite{berkooz_proper_1993, jolliffe_principal_2016, budanur_relative_2017}.

    We construct the dataset for PCA from long-time samples of the split edge, one-puff, and two-puff states. PCA yields an orthonormal basis ${\hat{Y}_i}$ of directions of maximal variation in the given data, ordered such that the captured variance decreases with increasing index. We use the first two components to construct the reduced phase space: projecting the instantaneous velocity field $\bm{u}(\bm{r},t)$ onto them defines a point in coefficient space $(p_1, p_2)$, where $p_i(t) = \int \mathrm{d}^3 x (\hat{Y}_i \cdot \bm{u})$.
    Then, each splitting event maps to a two-dimensional trajectory in this reduced space, starting near the one-puff and ending near the two-puff state. Note that the one-puff and two-puff states (as well as the split edge) do not reduce to a single point under the projection, and are instead represented by probability distributions (PDFs) in the $(p_1, p_2)$ plane.
    
\begin{figure*}[t] 
        \centering
        \includegraphics[width=2\columnwidth]{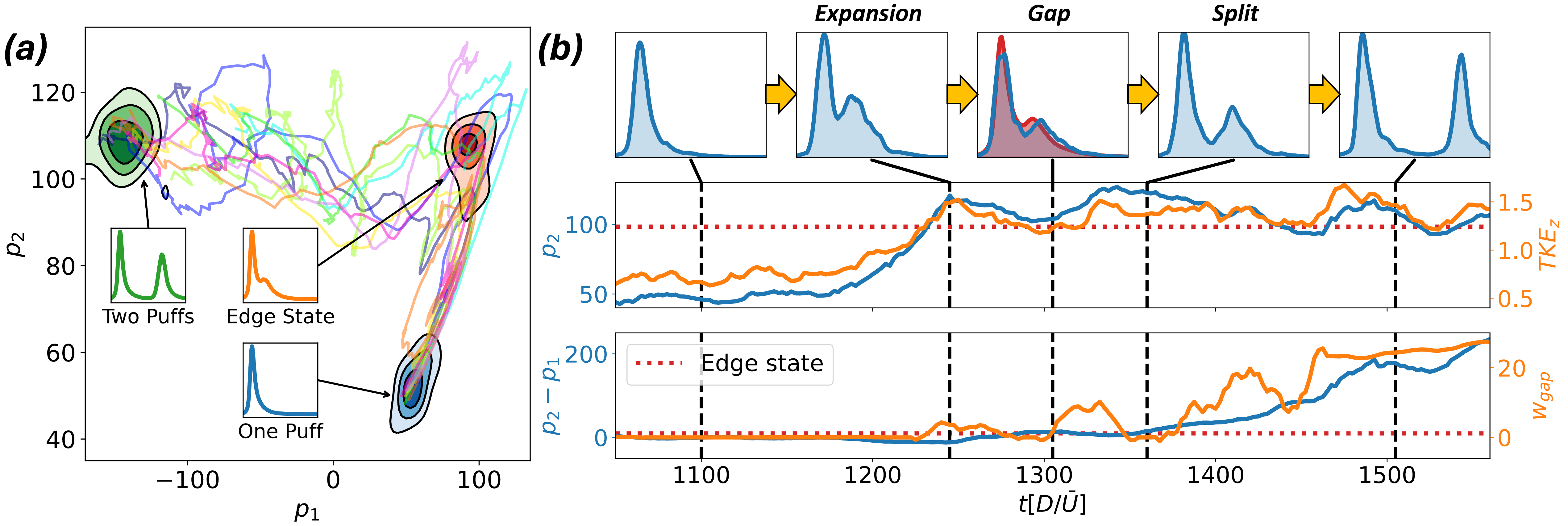}
        \caption{
         (a) Splitting trajectories in a reduced phase-space, projected onto the two largest PCA components, $p_1$ and $p_2$. PDFs of the puff (blue), split-edge state (orange), and two-puff states with a $25D$ gap (green) are shown with mean $q$ profiles (insets). Splitting experiments ($N=9$) appear as faint lines from the one-puff state, with initial segments (confined to the one-puff set) omitted. Puffs approach the edge by increasing $p_2$ (correlates with streamwise turbulent kinetic energy $TKE_z$) at nearly constant $p_2-p_1$, then split by increasing $p_2-p_1$ (correlates with gap width $w_{gap}$). (b) A single event: snapshots of $q$ (top), evolution of $p_2$ and $TKE_z$ (middle), and $p_2-p_1$ with $w_{gap}$ (bottom). Red dashed lines (middle, bottom) and the red $q$ profile (top) mark edge-state means.
         Puff splittings are consistent with the slug-gap-split scenario: (i) Puff expands into elongated puff with uniform turbulent core, approaching the ``split edge'' state; (ii) a laminar gap nucleates and grows until the split is complete.
        \label{fig:Re2200_spliting}}
    \end{figure*}

    Although trajectories vary, as expected in a high-dimensional chaotic system \cite{de_lozar_edge_2012, budanur_heteroclinic_2017}, they all pass near the split edge, forming a clear ``tube'' in the $(p_1, p_2)$ plane through which splits occur, Fig.~\ref{fig:Re2200_spliting}(a). In particular, on their way to the two-puff state, all trajectories pass near the split edge state, confirming its relevance. This is further supported by analyzing the $L^2$-distance between splitting trajectories and the split edge state, as described in the end matter section and shown in \cite{SI}, where separate figures for each trajectory are also provided.

\textbf{Puff-splitting mechanism}
    The reduced phase space representation can be used to reveal the puff-splitting mechanism at play. Examining the splitting trajectories in the $(p_1, p_2)$ plane, we find that they approximately follow two straight lines, corresponding to a two-step process. First, the trajectory travels along the line connecting the puff to the edge state, given by $p_1 \approx p_2$ and increasing $p_2$, and then it moves along the line of decreasing $p_1$ with $p_2$ roughly constant, connecting the edge state to the two-puff state. This implies that the relevant coordinates for the splitting mechanism are $p_2-p_1$ and $p_2$. However, since the $p_1, p_2$ observables were produced by PCA, they do not have a transparent physical meaning. We connect them to physically meaningful quantities by examining the correlations between $p_2$, $p_2 - p_1$ and various physical observables. We find significant correlations with two particular ones: The variable $p_2$ is strongly correlated with the turbulent kinetic energy (TKE) of the axial velocity, $\TKE_z$, defined as:
    \begin{equation}
        \TKE_z = \frac{1}{2} \int_V \dd^3 x \, (\bm{u} \cdot \hat{z})^2,
    \end{equation}
    and $p_2 - p_1$ is correlated with the laminar gap width between turbulent patches, $w_{\text{gap}}$, with correlation coefficients of $0.96$ and $0.91$, respectively (See \cite{SI} for the full correlation analysis). 
    
    This paints a compelling picture of a generic two-step process for puff splits as demonstrated for a specific split in Fig.~\ref{fig:Re2200_spliting}(b): First, as $\TKE_z$ increases, a turbulent core is formed inside the puff as it expands, resembling a short slug. At the end of this stage, the structure contains roughly the same TKE as the fully developed two-puff state, and matches the split-edge state, as shown by the red outline in Fig.~\ref{fig:Re2200_spliting}(b). Continuing along the transition, a gap forms inside the turbulent core, and the structure splits as this gap widens. It is instructive to note that $\TKE_z$ alone is not a sufficient indicator for splits, as only in approximately $50\%$ of cases where $\TKE_z$ reaches the edge-states' value does the structure split. In the other cases, the puff fails to split, as expected from an edge state.  
    
    The two-step process described above aligns with the previously proposed ``slug-gap-split'' mechanism \cite{frishman_mechanism_2022}: Rare fluctuations cause the puff to expand into a slug-like structure with a turbulent core. When wide enough, a minimal gap can nucleate within the core, allowing it to split into two puffs. The edge state corresponds to the ``tipping point'' between successful splitting and failure, which may result in retraction or the decay of a detached patch. We note that the precise distinction between splitting and non-splitting states remains unclear; we speculate that it depends on gap nucleation at specific locations, though a suitable observable has yet to be identified.
    
    Qualitative evidence for the slug-gap-split mechanism can be found in previous work on pipe flow \cite{shimizu_splitting_2014}, which can be reinterpreted in light of the proposed mechanism. Moreover, a similar mechanism seems to be at play for the self-replication of turbulence in plane Couette flow \cite{marensi_dynamics_2023}, indicating the possible universality of the slug-gap-split mechanism across different flow configurations.
     
\textbf{Exploring lower $\bm{\R}$}
    Although splitting events cannot be directly sampled at lower Reynolds numbers in DNS, the bisection algorithm can still be applied. At $\R = 2050$ and $\R = 2100$, the algorithm yields a semi-periodic edge state resembling a puff and a decay edge in relative motion - similar to the one found in the Barkley model for low enough $\R$ \cite{frishman_mechanism_2022}; see \cite{SI}. The observation of two different edge-states may reflect algorithmic limitations or a bifurcation near $2100 < \R < 2200$, with the slug-gap-split mechanism becoming relevant only at larger $\R$. Thus, the role of this periodic edge state remains unclear, and saddle avoidance \cite{borner_saddle_2024} may play a role here.

    Still, the slug-gap-split mechanism could be tested at lower $\R$ without identifying edge states or obtaining full flow fields. Instead, a state-space projection could be applied to splitting trajectories, using $\TKE_z$ and $w_{\text{gap}}$ to assess whether trajectories follow the predicted two-step path (as in Fig.~\ref{fig:Re2200_spliting}(a)). While DNS is limited by long waiting times \cite{avila_onset_2011}, experiments or rare-event methods like AMS \cite{gome_extreme_2022} could make this feasible.
    
\textbf{Conclusion}
    Our study reveals the dynamical structures and pathways underlying the transition to turbulence in pipe flow. It would be interesting to check the universality of the turbulence proliferation pathway found here, applying similar tools to other subcritical flows such as plane Couette or Taylor-Couette \cite{lemoult_turbulent_2014, manneville_transition_2016, klotz_phase_2022, marensi_dynamics_2023}, where sampling self-replication close to the critical point is more feasible.
    
    More broadly, we have demonstrated that an edge state on the boundary between two chaotic saddles can be successfully identified, and that it can play an important role in mediating transitions. To our knowledge, this is the first such study in a deterministic system with many degrees of freedom, demonstrating how the underlying transition mechanism can be revealed. Our approach should be applicable to general spatiotemporal chaotic systems, where transitions between coexisting chaotic saddles are of interest \cite{bouchet_rare_2019,borner_boundary_2025}. 

    Finally, it is noteworthy that although our system is deterministic, insights from stochastic models turn out to be extremely relevant. We find striking agreement with results from the Barkley model \cite{frishman_mechanism_2022, barkley_theoretical_2016}, highlighting the universality of the underlying dynamics and the models' ability to capture them minimally.

\bibliography{ref}

\section{End Matter}
\appendix
\subsection*{Direct numerical simulations of pipe flow}
    Considering the flow in a circular pipe as described in the main text,  the perturbation velocity is $\bm{u} = \tilde{\bm{u}} - \bm{u}_{\text{HP}}$, where $\tilde{\bm{u}}$ is the full flow velocity field and $\bm{U}_{\text{HP}}(r,\theta,z) = (1-r^2)\hat{z}$ is the laminar (Hagen-Poiseuille) in non-dimensional cylindrical coordinates ~\cite{acheson_elementary_1990}. The Navier-Stokes equations with boundary conditions for the perturbation velocity are then given by:
    \begin{equation}
    \label{eq:NS}
        \pdv{\bm{u}}{t} = -\bm{\nabla}p + 
            \frac{1}{\R} \nabla^2 \bm{u}  -
            \left(
                (\bm{U}_{\text{HP}} +\bm{u})\cdot\bm{\nabla}
            \right)
            (\bm{U}_{\text{HP}} +\bm{u}),
    \end{equation}
    together with the incompressibility $\bm{\nabla}\cdot\bm{u} = 0$ and a no-slip boundary condition at the wall $\bm{u}(1,\theta,z) = 0$. We integrate Eq.~\ref{eq:NS} using the \texttt{openpipeflow} pseudo-spectral code \cite{willis_openpipeflow_2017}. A spatial resolution of $N = 40$ radial points (at the roots of Chebyshev polynomials) was chosen, with Fourier modes up to $ \pm M = \pm 36 $ in the azimuthal direction ($\theta$) and up to $ \pm K = \pm 800 $ in the axial direction ($ z $) for a pipe length of $L = 150D$ (thus the total resolution is $1600\times72\times40$). The time step was fixed at $\Delta t = 2.5 \times 10^{-3} D/\bar{U}$, and a constant mass flux was enforced with $\bar{U} = 1$. 

\subsection*{Definition of the phase space boundary}
    We are interested in the phase-space boundary between initial conditions that reach the one-puff set ($S_{1}$) first and those that reach the two-puffs set ($S_{2}$) first. Similarly to the edge of chaos, a precise dentition requires considering a finite time horizon. We thus denote by $S_i^{\tau}$ $(i=1,2)$  the set of initial conditions that reach $S_i$ within $t \sim \tau$ and remain there for $t \gg \tau$. Here, $\tau$ should be chosen small enough relative to the split and decay timescales, yet large enough compared to the time-scale of fluctuations (since $S_i^{0} = S_i$, while for sufficiently large $\tau=T$,  $S_i^T = \varnothing$ as all puffs eventually decay or split). Then, for $\tau$ within this intermediate range, the sets $S_i^{\tau}$ should not depend on the precise value of $\tau$, making their definition meaningful.
    
    With this definition, $S_1^{\tau}$ and $S_2^{\tau}$ share a boundary consisting of states arbitrarily close to those in either set. The edge state is an attracting set on this boundary and thus can be found by following the dynamics along it, which should correspond to the stable manifold of this edge state. This is achieved by selecting two initial conditions $\bm{u}_1$ and $\bm{u}_2$ that are sufficiently close but ultimately evolve toward $S_1$ and $S_2$ after an initial transient (i.e. $\bm{u}_1 \in S_1^\tau$ and $\bm{u}_2\in S_2^\tau$).   
    
\subsection*{State classification}
    Applying the bisection algorithm requires a precise definition of the sets $S_1$ and $S_2$. This definition must distinguish between the two sets while remaining insensitive to variations within each set, motivating a coarse-grained approach.

    The $q$ variable (\eqref{eq:q_var}) is used to determine if a section of the pipe is turbulent or laminar. We define a threshold $q_{th}=5\times10^{-4}$, and classify the cross section turbulent (laminar) if $q \geq q_{th}$ ($q<q_{th}$). The chosen value for $q_{th}$ is consistent with commonly used ones, e.g. \cite{song_speed_2017}. 

    The one and two-puff states have distinct lengths and therefore one distinguishing feature is the total length of the pipe in a turbulent state. We define the total turbulent length as:
    \begin{equation}
        \label{eq:l_turb}
        l_{\text{turb}}(t) = \int_{0}^{L} \dd z \, \Theta(q(t,z) - q_{th})\,, 
    \end{equation}
    where $\Theta(x)$ is the Heaviside step function. Note that we use $l_{\text{turb}}$ rather than the more commonly used turbulent fraction $l_{\text{turb}}/L$ as we want the definitions to be independent of the pipe's axial length. In previous applications of the bisection algorithm (e.g. \cite{mellibovsky_transition_2009}) $l_{\text{turb}}$ alone would be sufficient to differentiate between the two sets. In those cases, a significant decrease in $l_{\text{turb}}$ reliably indicated relaminarization. Here, however, the challenge arises from the transient nature of both states under consideration. While, in principle, $l_{\text{turb}}$ would approach the characteristic value of either one or two puffs if integrated for a sufficiently long time, extending the integration window increases the risk of an additional state transition. As such, our goal is to classify the states as early as possible -- both to avoid misclassification and to ensure the algorithm remains computationally efficient. Simply observing an initial increase or decrease in $l_{\text{turb}}$ is not sufficient to reliably classify the state, as in many cases both escapes to $S_1$ and $S_2$ exhibit an initial increase in $l_{\text{turb}}$. Instead, to distinguish the two sets, it is essential to consider the presence of a laminar gap in the $q$ profile for states in $S_2$ (corresponding to the laminar flow between two puffs) and its absence in states belonging to $S_1$. The necessity of imposing such an additional classification constraint has also been observed in the channel flow setting~\cite{gome_extreme_2022}.
    
    Thus, the one-puff and two-puff states are characterized based on three criteria: the total length of the pipe in a turbulent state $l_{\text{turb}}$, the number of distinct turbulent patches $n(t)$, and the width of the laminar gaps between them $w_{\text{gap}}$, where $n(t)$ is given by the number of turbulent patches separated by laminar gaps, and $w_{\text{gap}}$ is defined as the smallest laminar region (along $z$) between those patches.
    
    To find suitable values for the above parameters for each set, we perform two long experiments for each $\R$ considered: one with a single puff and another with two puffs separated by a minimal distance of approximately $25 D$. We measure the mean $\Bar{l}_i$ and standard deviation $\sigma_i$ of $l_{\text{turb}}(t)$ in the case of one ($i=1$) and two ($i=2$) puffs and $w_{\text{gap}}$ for the two-puff state. Statistics are gathered over the time of $1000 D/\bar{U}$. 
    
    We define a classification function of the instantaneous flow state $\Phi[\bm{u}(t)]$ such that $\Phi = 1$ corresponds to an instantaneous single puff, $\Phi = 2$ to two puffs, and the flow state is considered transitional for $\Phi = 0$. $\Phi=-1$ indicates all undesirable states, such as laminar and three puffs. With the above, we take the classification function to be:
    \begin{equation}
    \label{eq:Phi}
        \Phi = 
            \begin{cases}
            2 
            & \text{: } \abs{l_{\text{turb}} - \Bar{l}_2} <  2\sigma_2
            \text{ and } n = 2 \text{ and } w_{\text{gap}} > w_{th}\\
            1  
            & \text{: } \abs{l_{\text{turb}} - \Bar{l}_1} <  2\sigma_1
            \text{ and } n = 1\\
            -1
            & \text{: } l_{\text{turb}} > \Bar{l}_{1} + \Bar{l}_{2} 
            \text{ or } l_{\text{turb}} < 1\\
            0
            & \text{: otherwise}
            
            \end{cases}
    \end{equation}
    Because both of the considered states are transient, a state is classified as a puff (two-puffs) if $\Phi(t) = 1 $ ($\Phi(t) = 2 $) is constant for a time of at least $\tau = 50 D/\bar{U}$. In the case that $\Phi(t) = -1$ is ever encountered the procedure is terminated (which never occurred in our investigation).
    
    The parameters used:
    For $\R= 2200$: $\Bar{l}_1 = 10.75$, $\sigma_1= 2.3$, $\Bar{l}_2 = 21.5$, $\sigma_2 = 3.1$. 
    For $\R= 2100$: $\Bar{l}_1 = 9.4$, $\sigma_1= 2.1$, $\Bar{l}_2 = 18.8$, $\sigma_2 = 2.5$.
    For $\R= 2050$: $\Bar{l}_1 = 9.05$, $\sigma_1= 2$, $\Bar{l}_2 = 18.1$, $\sigma_2 = 2.4$. And $w_{th} = 20$ for all $\R$.

\subsection*{Implementation of edge tracking algorithm}
    The bisection algorithm for edge-tracking follows the procedure described in \cite{skufca_edge_2006}. To implement this procedure, we require three components: a classification method to distinguish between states, a meaningful metric to define a measure of closeness, and a method to refine initial conditions with respect to said metric.

    The classification of states is based on the observable defined in \eqref{eq:Phi}. Starting from a given flow state $\bm{u}(t_0)$, the system is integrated forward in time until $\Phi(t) \equiv \Phi[\bm{u}(t > t_0)]$ stabilizes at $\Phi = 1$ (or $\Phi = 2$) for a duration $\tau$, after which the state is assigned to $S_1^\tau$ (or $S_2^\tau$).
    
    Next, we define a metric to quantify the distance between states. We require that states related by a symmetry of the pipe yield a distance of zero ~\cite{willis_revealing_2013}. Accordingly, we employ a translational symmetry-reduced $L^2$ norm, defined as
    \begin{equation}
        \label{eq:dist}
        d(\bm{u}_a, \bm{u}_b)  
        = \min_{\zeta} \left[  
        \int \dd^3 x \abs{
        \bm{u}_a(r,\theta, z) - \bm{u}_b(r, \theta, z - \zeta)  
        }^2  
        \right]^{1/2}.
    \end{equation}  
    In principle, the metric we use should be invariant under the symmetries of the system, so rotations along $\theta$ should also be considered in the minimization. However, we find that they do not significantly affect the distance measure beyond the initial transient phase of the algorithm, as the bounding states exhibit no notable azimuthal rotation over the relevant time-scales. This metric is also used to find the maximal Lyapunov exponent, as stated in the main text.

    Finally, for the refinement step, we adopt the most common approach: given states $\bm{u}_a$ and $\bm{u}_b$, we generate a new flow state $\bm{u}_c$ by bisecting the chord connecting them via linear interpolation. This yields $d(\bm{u}_c, \bm{u}_a) = d(\bm{u}_c, \bm{u}_b) < d(\bm{u}_a, \bm{u}_b)$.

    With the above definitions, the algorithm proceeds as follows: each bisection step begins with two bounding states, $\bm{u}_1\in S_1^\tau$ (one puff) and $\bm{u}_2\in S_2^\tau$ (two puffs). A new flow state, $\bm{u}_0$, is generated by linearly interpolating between the two bounding states. To classify this new state, it is integrated forward in time until the observable $\Phi(t)$ stabilizes at either $\Phi = 1$ or $\Phi = 2$ for a duration of $\tau = 50 D/\bar{U}$. Based on this result, the appropriate bounding state is updated to $\bm{u}_0$, and the bisection step is repeated until $d(\bm{u}_1, \bm{u}_2) < h_1 = 2 \cdot 10^{-6}$ and both of the original bounding states have been replaced. The latter condition ensures that the bounding state is always neither strictly in the one-puff nor the two-puff set (i.e., it belongs to $S_i^\tau$ but not to the set $S_i$), and initially satisfies $\Phi(t = 0) = 0$.

    Next, both bounding states, $\bm{u}_1 $ and $\bm{u}_2$, are integrated forward in time until their separation reaches $d(\bm{u}_1(T), \bm{u}_2(T)) = h_2 = 5 h_1$, but for no longer than $25 D/\bar{U}$. During this forward integration, the bounding flow fields are sampled every $200 \Delta t$ for later analysis. The entire procedure is then repeated with the updated bounding states, $\bm{u}_i \leftarrow \bm{u}_i(T)$ ($i=1,2$), where $T$ is the separation time. Over long times, the pair $(\bm u_1(t), \bm u_2(t))$ consequently tracks the boundary until it finds the edge state. The robustness of the results was verified by varying $10^{-7} < h_1 < 10^{-3}$ and $25 < \tau \bar{U} / D < 100$, without any noticeable variation in the results.
    
    \subsection*{Principal Component Analysis}
    The dataset we have used for the PCA consists of an equal number of snapshots (equally spaced in time) of the split edge, single puff, and two-puff states, using 1000 snapshots for each. 
    In addition, we employ symmetry reduction for each snapshot \cite{willis_revealing_2013, budanur_relative_2017}, azimuthally averaging and axially shifting the flow field so that the upstream front is at $z=0$. Such a reduction is necessary since otherwise states related by a simple coordinate shift correspond to directions of large variation. Furthermore, because we are interested in the two-puff state immediately after splitting, and different spacings between the two puffs correspond to different velocity fields even with symmetry reduction, we select as a representative two-puff state states with a gap of $w_{\text{gap}} = 25 \pm 5 [D]$. 

    In presenting the splitting trajectories, we have used the first two principal components. Since the dataset is very high-dimensional, two components naturally do not capture the entire variation in the data. To ensure that the PCA representation is not misleading, we validated the transition picture presented in the main text using an alternative method. Specifically, we have assessed the proximity of each splitting trajectory to the edge state using the symmetry-reduced snapshots described above both for the splitting experiments and the split edge state. 
    For each splitting experiment $\bm{u}(t)$, we measured the distance $d_{es}(t) \equiv d(\bm{u}(t), \bar{\bm{u}}_{es})$, where $\bar{\bm{u}}_{es} =\langle \bm{u}_{es}(t)\rangle_t$ is the time-averaged split edge-state, and $d$ is given by \eqref{eq:dist}. To quantify proximity, we defined a distance scale $\delta_{es}$ as the expected distance between edge-state snapshots and their mean, $\delta_{es} = \langle d(\bm{u}_{es}(t), \bar{\bm{u}}_{es}) \rangle_t$.
    We find that, for all trajectories considered,  $d_{es}(t)$  attains this minimum, $\min_t d_{es}(t) \lesssim \delta_{es}$, during the transition. Moreover, this minimum occurs at the same time the trajectory is closest to the split edge state in the $(p_1, p_2)$ representation. This provides additional validation of the accuracy of the PCA representation. See \cite{SI} for further details.

\end{document}